\documentclass{article}
\usepackage{graphicx} 
\usepackage{url,hyperref,lineno,microtype,subcaption}
\usepackage[onehalfspacing]{setspace}
\usepackage{amsmath, amssymb }
\usepackage{cite}

\newcommand{\sign}{\text{sign}}
\usepackage{authblk}

\title{Quantum-inspired algorithm for simulating viral response}

\author[1]{Daria O. Konina}
\author[1]{Dmitry I. Korbashov}
\author[1]{Ilya V. Kovalchuk}
\author[1]{Aygul A. Nizamieva}
\author[1,2]{Dmitry A. Chermoshentsev}
\author[1]{Aleksey K. Fedorov}

\affil[1]{Russian Quantum Center, 30 Bolshoy Boulevard, building 1, Skolkovo Innovation Center territory, Moscow, 121205, Russia}
\affil[2]{Skolkovo Institute of Science and Technology, 30 Bolshoy Boulevard, building 1, Skolkovo Innovation Center territory, Moscow, 121205, Russia}

\date{January 2025}

\begin{document}

\maketitle
\begin{abstract}
Understanding the properties of biological systems is an exciting avenue for applying advanced approaches to solving corresponding computational tasks. A specific class of problems that arises in the resolution of biological challenges is optimization. In this work, we present the results of a proof-of-concept study that applies a quantum-inspired optimization algorithm to simulate a viral response. We formulate an Ising-type model to describe the patterns of gene activity in host responses. Reducing the problem to the Ising form allows the use of available quantum and quantum-inspired optimization tools. We demonstrate the application of a quantum-inspired optimization algorithm to this problem. Our study paves the way for exploring the full potential of quantum and quantum-inspired optimization tools in biological applications.

\end{abstract}

\section{Introduction}
Viruses have been existing alongside humankind throughout history. They are generally characterized by great genetic diversity and their ability to change and conform. So far, viruses have been a powerful evolutionary tool for regulating the size and viability of populations. For example, it has been proven that retroviruses and retroelements have a long-term impact on evolution, creating new genes and increasing species genome complexity \cite{Svoboda2011}. Every year humanity faces the rise of new types of viruses, some causing severe pandemics, such as the recent example of the SARS-CoV-2 virus \cite{Flerlage2021, Sironi2020}. It is essential to understand the underlying molecular and genetic pathways of the immune response to virus exposure, as it directly affects the propagation of the disease. Identifying the molecular signatures critical to a biological process of viral response is crucial to the development of new possible treatment methods. 

With diverse and complex biological data, practical assessment methods for their processing are highly demanded. In recent studies, new approaches, such as machine learning or hypergraphs, have revealed new key elements of biological processes. Special-purpose physical devices, known as coherent Ising machines, which can effectively solve combinatorial optimization problems, are actively investigated and applied to different practical problems~\cite{Inagaki2016, Mcmahon2016, Yamamoto2017, Hamerly2019, Okawachi2020, Reifenstein2021}. The theoretical investigation and simulation of such devices have become a basis for developing a new generation of optimization algorithms, which are referred to as ``quantum-inspired" tools. 

In this work, we evaluate the applicability of the quantum-inspired algorithm SimCIM~\cite{Tiunov2019}, which simulates the operation of an optoelectronic coherent Ising machine (CIM), to solve a problem of complex biological system modeling. 
Such a technique is promising for finding practical low-energy solutions for problems that are intractable for classical computers across disciplines. 
Developing new methods for quantum hardware is an important task that opens up new horizons in future research.

This paper presents a proof-of-concept study of the application of the SimCIM algorithm to simulate the viral response. We consider a set of viruses and, with available experimental data on viral response in host cells, we transfer the data to a form of an Ising model to establish the patterns of gene activity in host responses.

\section{Materials and Methods}

For our study, the seven Microarray data about different viral infections containing GSE80059, GSE76600, GSE80697, GSE45399, GSE79458, GSE86528, GSE33267, and GSE48142 were obtained from the Gene Expression Omnibus (GEO) repository \cite{Omnibus}. The main features of these seven datasets are shown in Table \ref{tab1}. 

\newcommand{\specialcell}[2][c]{%
  \begin{tabular}[#1]{@{}c@{}}#2\end{tabular}}

\begin{table}
\caption{The main features of  selected datasets.}
\begin{center}
\scalebox{0.8}{
\begin{tabular}{|c|c|c|c|c|}
\hline
GEO Datasets & Virus infection & \specialcell{ \\ Target RNA class \\ in  Microarray} & Cell Type & Time points\\[20pt]
\hline
GSE80059 & Ebola virus & miRNA & Huh & \specialcell{\\ 8, 18, 24, \\ and 36h}\\[20pt]
\hline
GSE76600 & Influenza Virus & miRNA & Calu-3 & \specialcell{ \\ 0, 7, 12, \\ and 24 h} \\[20pt]
\hline
GSE80697 & Influenza Virus & mRNA & Calu-3 & \specialcell{\\ 0, 7, 12, 24, 36, \\ and 48 h}\\[20pt]
\hline
\specialcell{\\ GSE45399} & \specialcell{\\ Influenza Virus} & \specialcell{\\ lncRNA} & \specialcell{\\ A549} & \specialcell{\\ 8 and 24h}\\[20pt]
\hline
GSE79458 & MERS-coronavirus & mRNA & Human fibroblasts & \specialcell{\\ 0, 12, 24, 36, \\ and 48 h}\\[20pt]
\hline
GSE86528 & MERS-coronavirus & mRNA & Human fibroblasts & \specialcell{\\ 0, 12, 24, 36, \\ and 48 h}\\[20pt]
\hline
GSE48142 & SARS-coronavirus & mRNA & 2B-4 & \specialcell{\\ 0, 7, 12, 24, 36,  \\ 36, 48, 60 or 72h}\\[20pt]
\hline
\end{tabular}\label{tab1}
}
\end{center}
\end{table}

All raw data from GSE80059, GSE76600, GSE80697, GSE45399, GSE79458, GSE86528 and GSE48142 were merged using the Python GEOparse v2.0.3 package \cite{Geoparse}.

\subsection{Identification of Differentially expressed genes (DEGs)}

For the identification of differentially expressed genes (DEGs) in the chosen datasets, background correction and normalization were performed. The datasets have been processed for further analysis using the web-based analytical tool GEO2R \cite{Geo2r}. GEO2R employs built-in linear models for microarray data (limma) from the R package and GEOquery. The R package facilitated the computation of $|\log_2 FC|$ values and the false discovery rate (FDR)-adjusted values for each gene. Default parameters were applied during the preprocessing of the datasets.

Furthermore, the GEOexplorer Web server \cite{Geoexplorer} was utilized to investigate the molecular basis of phenotypic and biological differences between groups of biological samples and to compare datasets of interest \cite{Hunt2022}. Differentially expressed genes were extracted and classified according to their biotypes: mRNA, lncRNA, and miRNA (p-value: $p < 0.05$).

Pathway annotation and enrichment analyzes for the identified DEGs have been performed using the Gene Ontology (GO) and Kyoto Encyclopedia of Genes and Genomes (KEGG) pathway databases. Gene Set Enrichment Analysis (GSEA) was also performed to determine whether the identified DEGs (specifically for the mRNA biotype) exhibited significant and consistent differences between two biological states.

\subsection{Topological Overlapping Matrix (TOM) generation}\label{section_2.2}

A typical result of microarray analysis is a list of genes (or their products) and their expression levels under specific conditions or at a particular time. Network-based approaches, such as coexpression networks, are often employed for further analysis to uncover complex interrelationships between molecules and their functions within the experimental system under study.

In our work, we have used the PyWGCNA Python library \cite{Rezaie2022} to reconstruct the coexpression network of the treated and control groups. In the first step of the WGCNA method, the gene expression matrix was converted into a pairwise gene similarity matrix using the Pearson correlation test. The similarity matrix was then transformed into an adjacency matrix. Following the standard methods in PyWGCNA, we further converted the adjacency matrix into a topological overlap matrix (TOM).

\subsection{Identification of Key Genes associated with Viral Response (IKGVR) problem}

When a virus infects a host, the immune system responds by initiating a cascade of complex biological reactions. In this study, we use a molecular interaction network approach to investigate host response processes, including the activation of various immune cells and the release of various signaling molecules. These complex biological processes involve a set of genes participating in multi-way interactions across distinct reaction cascades.

To identify key regulatory genes with maximum possible multiway interactions that play a central role in the host viral response, we have decided to explore graph-based solutions. Graphs are frequently used to model these interactions \cite{Taheri2023}, \cite{Renaux2023}, \cite{Karami2021}, \cite{Sci_Li2018}. For our purpose, our goal is to develop a graph that highlights clusters or modules of genes that are tightly interconnected, potentially revealing functional gene groups or pathways.

Firstly, we construct the TOM of the expression datasets, as described in Section \ref{section_2.2}. The topological overlap quantifies the similarity between two genes based on their coexpression relationships with all other genes, resulting in a final gene coexpression score matrix. Each cell in the matrix represents the topological overlap value between two genes, with values closer to 1 indicating stronger similarity or interaction.

After constructing the TOM matrices, we applied a threshold to filter the TOM values and define interactions between genes. Values above the threshold are considered significant, indicating a connection between the corresponding genes, while values below the threshold are discarded, implying weak or no interaction. In the resulting graph, nodes represent genes, and edges represent significant interactions, as determined by the chosen threshold. Based on the distribution of TOM values and gene correlation between co-expressed modules, we selected a threshold of 0.3~\cite{TAN201158,Chang2023}. Therefore, a graph $G$ is created from the TOM matrix, where vertices correspond to genes and vertices are connected by an edge if  $TOM[gene A][gene B] > 0.3$. 

Using the graph $G$, we aimed to identify a set of key genes, with the set size determined to be 15, consistent with similar studies~\cite{Taheri2023,Virus_Chen2022SARS_Chen2022}. To simplify the problem, we selected a subset of vertices with the maximal degrees (size $n_0$ = 100 for the subset was also chosen based on the published data~\cite{Taheri2023,Virus_Chen2022,SARS_Chen2022,Feng2021}, and the best 15 vertices with high connectivity were chosen from them.

%

Taking into account the above, we can formulate the problem as follows. Given a graph $G = (V,E)$,  where $|V| = n$ ($n$ is the number of genes in the TOM matrix) and the subset $M\subseteq V$ such that $|M|$ = $n_0$ = 100 ($n_0$ is the subset of $V$ vertices with maximum degrees), our objective is to find a subset $U \subseteq M$ such that $|U| = k = 15$. 

This leads to the following mathematical formulation:
\begin{equation}\label{graph_problem}
\left| \bigcup_{u \in U} \{ v \in V : (u, v) \in E \} \right| \to  max
\end{equation}
where  $M\subseteq V$ in graph $G = (V,E)$: \begin{equation}
M = \{ v_1,\dots,v_{n_0}\}: \max\left(\sum_{i=1}^{n_0=100}deg(v_i)\right), v\in V
\end{equation}

\begin{figure}[h]
\centering
\includegraphics[width=1\linewidth]{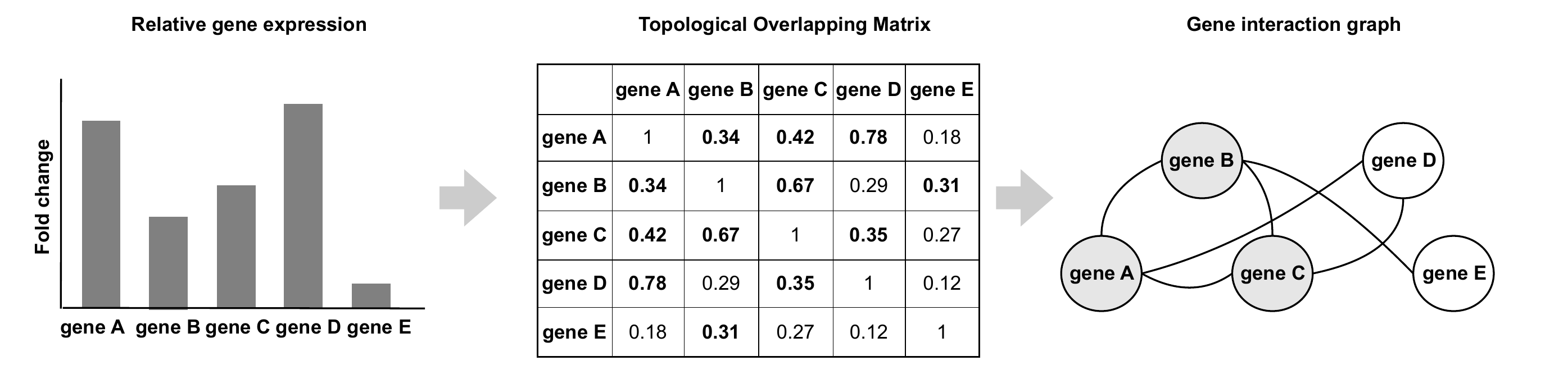}
\caption{Problem statement example. (Left) - relative gene expression for 5 genes. (Center) - TOM for genes expression data. (Right) - visualization of a corresponding graph. The selected key genes are highlighted in gray - the set of vertices with the maximum degrees is indicated.}
\label{fig:mpr0}
\end{figure}

The problem of finding a subset $U\subseteq M$ such that $|U| = k$ and the number of vertices in $G$  adjacent to $U$ is maximized is NP-complete. This is because it is a special case of the maximum coverage problem and a generalization of the dominating set problem \cite{Lucas2014},\cite{Garey1979}.

NP-complete problems are challenging because of their exponential growth in complexity as the size of the problem increases. In our work, we employ the quadratic unconstrained binary optimization (QUBO) method to address this problem.


\subsection{Transforming IKGVR problem to a QUBO form}


The QUBO problem is formulated as follows.
Find the global minimum point of a pseudo-Boolean ``energy" function $E: \mathbf{q}  =\{0, 1\}^N \rightarrow\mathbb{R}$ such that: 
		\begin{equation}\label{equ:pubo}
			E(\mathbf{q}) = \mathbf{q}^T \mathbf{A} \mathbf{q} + const,
		\end{equation}
where $N$ is the dimension of a problem instance (problem size), and $A$ is a square matrix of size $N$ with real coefficients.

 To transform the IKGVR problem (\ref{graph_problem}) into a QUBO formalism, we follow several steps. In the first step, we introduce $n_0$ variables $q = \{ q_1,\dots,q_{n_0}\}$, with the following encoding: 
 \begin{equation}\label{qubo_coding}
 q_{i}=\begin{cases} 1 & \mbox{for}\ v_i \in U, \\
			\ 0 &\mbox{otherwise}.
    \end{cases}
\end{equation}

 
Since our goal is to find a subset $U\subseteq M$ such that $|U| = k$, the following constraint ($E_p$) should be included in the optimization function:
\begin{equation}\label{qubo_init}
   E_{p}(\mathbf{q}) = \lambda \cdot\left(\sum_{\substack{0<i\leq n_0}} q_i - k\right)^2, 
 \end{equation}
where $\lambda$ is the penalty coefficient for solutions that do not satisfy the condition $|U| = k$. 
Thus, we have established a bijection between these variables and the various solutions to the problem.
 
Let $N(v, U) := |N_G(v) \cap U|$, where $N_G(v)$ is the set of neighbors to which vertex $v$ is connected in $U$. Therefore, $N(v, U)$ is linear in the variables $\mathbf{q}$ and can be written as follows:
\begin{equation}\label{n_vU}
 N(v, U) = \sum_{\substack{0<i\leq n_0}} q_i \cdot A(G)_{vi},
 \end{equation}
where $A(G)$ is the adjacency matrix of the graph $G$. 
 
 Taking into account the variables notations (\ref{n_vU}), the maximization problem (\ref{graph_problem}) is equivalent to maximizing the following function:
 \begin{equation}\label{max_n_vU} 
 \sum_{\substack{v \subseteq G}} \min(N(v, U), 1) = \sum_{\substack{v \subseteq G}}\min \left(\sum_{\substack{ 0<i\leq n_0}} q_i \cdot A(G)_{vi}, 1\right) \to  max 
  \end{equation}

To transform the function (\ref{max_n_vU}) into QUBO form, we add $\lceil \log_2(n_v) - 1 \rceil$ additional variables $\mathbf{b}^{(v)}$ for each vertex $v$ in $G = (V,E)$ such that $n_v > 2$, where $n_v := |N_G(v) \cap M|$. The meanings of $\mathbf{b}^{(v)}$ are defined in the optimization process. Finally, the following term is obtained:
 
 \begin{equation}\label{summ}
 E^{(v)}(\mathbf{q}, \mathbf{b}^{(v)}) = \frac{1}{2}\left( N(v, U) - \sum_{\substack{0<i\leq \lceil \log_2(n_v)-1 \rceil
 }} 2^i \cdot  b^{(v)}_i - \frac{3}{2}\right)^2, 
 \end{equation}
 there:
 \begin{equation}\label{b_i}
 \left(\sum_{\substack{0<i\leq \lceil \log_2(n_v)-1 \rceil}} 2^i \cdot  b_i\right) \in \mathbb{Z}_{\ge0} \in [0;2^{\lceil\log_2(n_v)-1 \rceil + 1})
 \end{equation}

 and the following inequality is satisfied:
 \begin{equation}
 2^{\lceil \log_2(n_v)-1 \rceil + 1}\geq n_v \geq N(v, U)
 \end{equation}
 
 At optimal points, it is necessary for the function \ref{summ} to be close to zero. Hence, if $N(v, U) > 0$, the function (\ref{summ}) takes $\frac{1}{2} \cdot  0.5^2 = 0.125$; otherwise, it takes the value $\frac{1}{2} \cdot  1.5^2 = 1.125$. Thus, the term of the function (\ref{summ}) actually maximizes the function (\ref{max_n_vU}).

The function for $E^{(v)}(\mathbf{q}, \mathbf{b}^{(v)})$ is presented as follows:
 \begin{equation}\label{summ_with_const}
 E^{(v)}(\mathbf{q}, \mathbf{b}^{(v)}) = \frac{1}{2}\left(\sum_{\substack{ 
  0<u\leq n_0
 }} q_i \cdot A(G)_{vi} - \sum_{\substack{ 
  0<i\leq \lceil \log_2(n_v)-1 \rceil
 }} 2^i \cdot  b^{(v)}_i - \frac{3}{2}\right)^2, 
 \end{equation}
 
Finally, the resulting optimization function takes the following form:
\begin{equation}\label{qubo_form}
 E_{tot}(\mathbf{q}, \mathbf{b}) = \sum_{\substack{0<v\leq n}} E^{(v)}(\mathbf{q}, \mathbf{b}^{(v)}) + E_p(\mathbf{q})
\end{equation}
here $\mathbf{b}$ is a vector that included all the ancilla variables for all $n$ vertices. 

\subsection{Quantum-inspired optimization via SimCIM algorithm}
 
SimCIM is an algorithm for solving QUBO problems by simulating CIM devices~\cite{Tiunov2019}. To solve optimization problems using quantum annealers, such as CIM devices, we need to transform the QUBO problem \ref{qubo_form} into an Ising model. 
Transforming a QUBO problem into an Ising model is a standard process that allows optimization problems formulated in terms of binary variables $(q_i \in {0, 1})$ to be expressed in terms of spin variables $(s_i \in {-1, 1})$. The relationship between the binary variables $q_i$ and the spin variables is given by: $s_i := 2 \cdot q_i - 1$. The Ising model energy function can be written as follows:
\begin{equation}\label{ising_form}
 E(\mathbf{s}) = \sum_{\substack{<i,j>}} J_{ij}s_is_j + \sum_{\substack{i}}h_is_i
\end{equation} 
where $J_{ij}$ is the spin–spin interaction, $h_i$ is the local field applied to each spin.
 
Physically, each CIM consists of an optical parametric oscillator formed by a degenerate parametric amplifier (a single-mode squeezer) inside a fiber loop, with $N$ optical pulses passing through the loop. The continuous values of the position quadrature of the pulse modes correspond to the $N$ variables. Iteratively, the quadrature of each pulse is measured and subjected to a phase-space shift along the position axis. This shift is calculated from the gradient of the optimization function, after which the pulses are amplified to a coherent state of a particular intensity. 
As a result, the pulse phases are stabilized at either $\pi$ or $0$, and the final sequence of $s_i = \pm 1$, corresponding to the solution of the Ising problem, is obtained.

The algorithm starts with a zero vector of continuous quadratures and then iteratively calculates the displacement $\Delta s_{i}(t)$, adding it to the vector of quadratures. The iteration of the SimCIM algorithm is represented as: \begin{equation} \Phi_{i} = \sum_{j} J_{ij} s_{j} + h_i \end{equation} The displacement, which is equivalent to the gradients for the spin values, is calculated as follows:
\begin{equation}\label{equ:displacement}
		\Delta s_{i}(t) = \nu(t) s_{i} + \xi \left(\sum_{j} J_{{i}{j}}s_{j} + h_i\right) + f_{i}(t)
\end{equation}
Here, $\nu(t)$ accounts for the combination of the time-dependent pump amplitude and linear loss, $\xi$ is equivalent to the learning rate in the conventional gradient descent algorithm, and $f_i$ represents Gaussian noise with variance $\sigma^2$. We use the momentum method \cite{Qian1999} to accelerate convergence, with the momentum parameter $\alpha$.

An activation threshold function, $\phi(s_i): s_i \gets \phi(s_i + \Delta s_i)$, is applied to each variable to account for saturation:
\begin{equation}\label{equ:saturation}
    \phi(s_{i})=\begin{cases} s_{i} & \mbox{for}\ |s_{i}| < s_{\rm sat}, \\
			\sign(s_i) \cdot  s_{\rm sat} &\mbox{otherwise}.
    \end{cases}
\end{equation}

In the above equations, $\nu(t)$, $\xi$, $\sigma$, and $s_{\rm sat}$ are hyperparameters of the algorithm that require fine-tuning for each problem instance. 
In the original paper~\cite{Tiunov2019}, $\nu(t)$ grows with time according to the hyperbolic tangent law. After the last iteration, the obtained ``solution'' is evaluated as $\sign(s_i)$.

\subsection{ceRNA network analysis for SimCIM results}

In our work, we analyzed microarray data for three RNA biotypes. To explore and verify genes identified through quantum algorithms, we constructed a regulatory network of competitive endogenous RNAs (ceRNAs) based on the ceRNA hypothesis \cite{Salmena2011}. In the ceRNA network, we aim to understand possible mRNA-miRNA and mRNA-lncRNA interactions. We used various online tools and databases to identify and predict target relationships. Targets of miRNAs were identified using Mirnet \cite{Chang2020} (https://www.mirnet.ca/), while the RISE database of RNA interactomes from sequencing experiments \cite{Gong2018} was used to predict lncRNA–RNA interactions.


\section{Results}
\subsection{Preprocessing and DEG Analysis}
\begin{figure}[h!]
\centering
\includegraphics[width=1\linewidth]{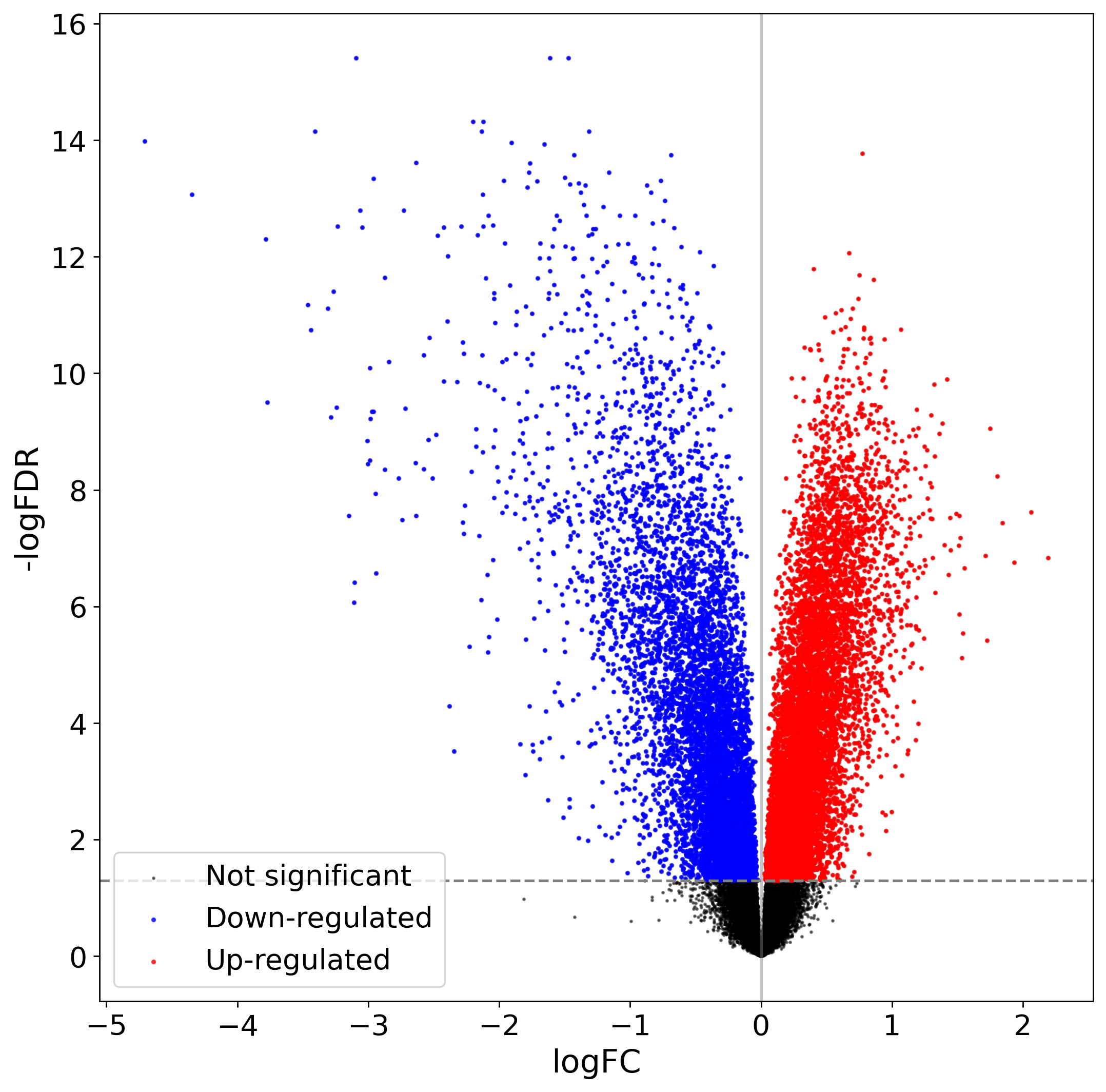}
\caption{Volcano plot visualization of DEGs (p-value $<$ 0.05) that were identified by analysis expression differences in two groups: Group 1(mock-treated cells) and Group 2(virus-treated cells).
Red dots mean up-regulated genes, blue dots mean down-regulated genes, black dots mean not differential expressed. The presented results were calculated for the GSE80697 dataset.}
\label{fig:mpr1}
\end{figure}

\begin{figure}[h]
\centering
\includegraphics[width=1\linewidth]{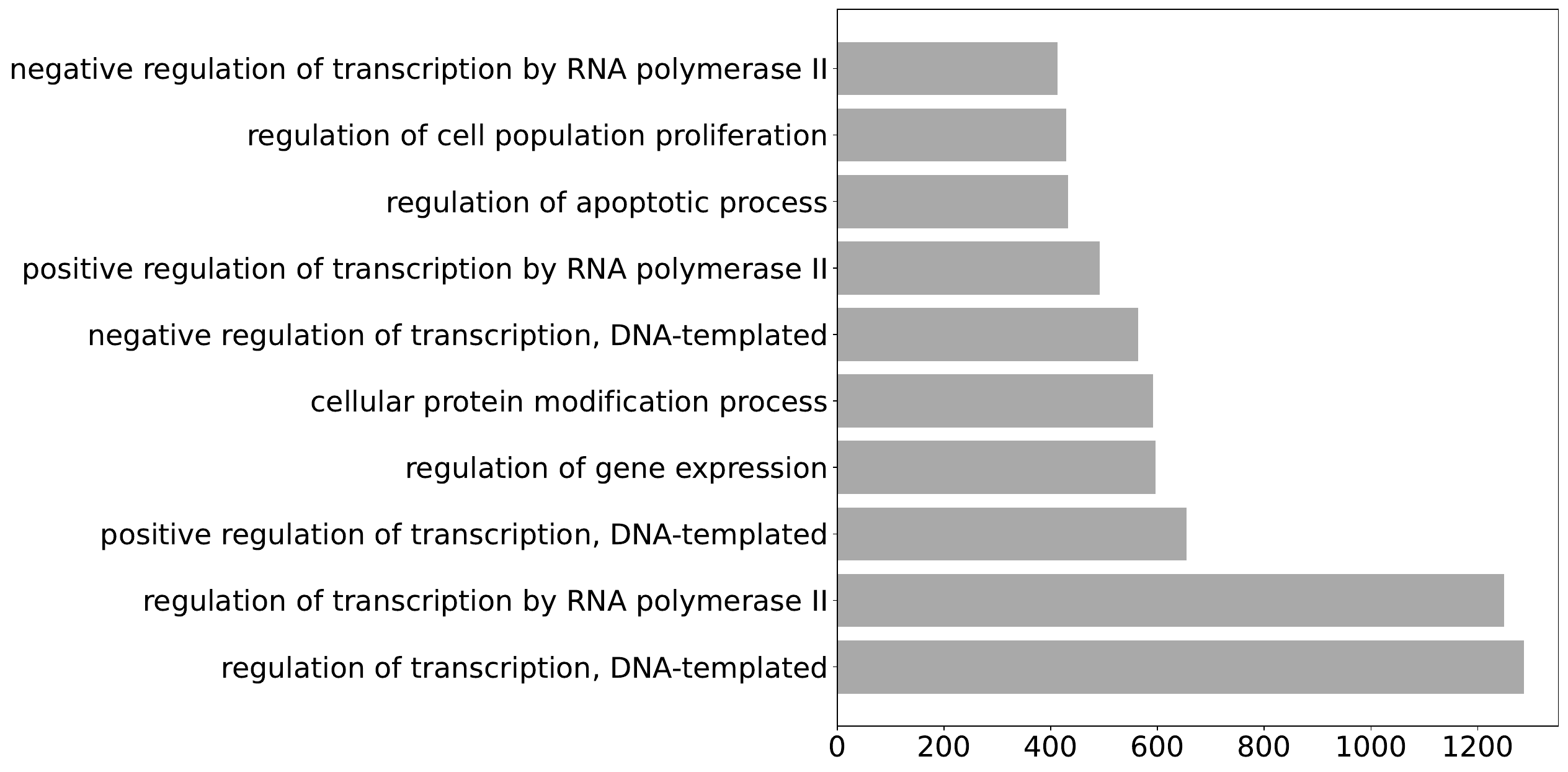}
\caption{Resuls of GO pathway analysis of all up- and down-regulated genes in host viral response, visualization of 10 most common GO pathways for DEGs. The presented results were calculated for the GSE80697 dataset.}
\label{fig:mpr2}
\end{figure}

\begin{figure}[h]
\centering
\includegraphics[width=1\linewidth]{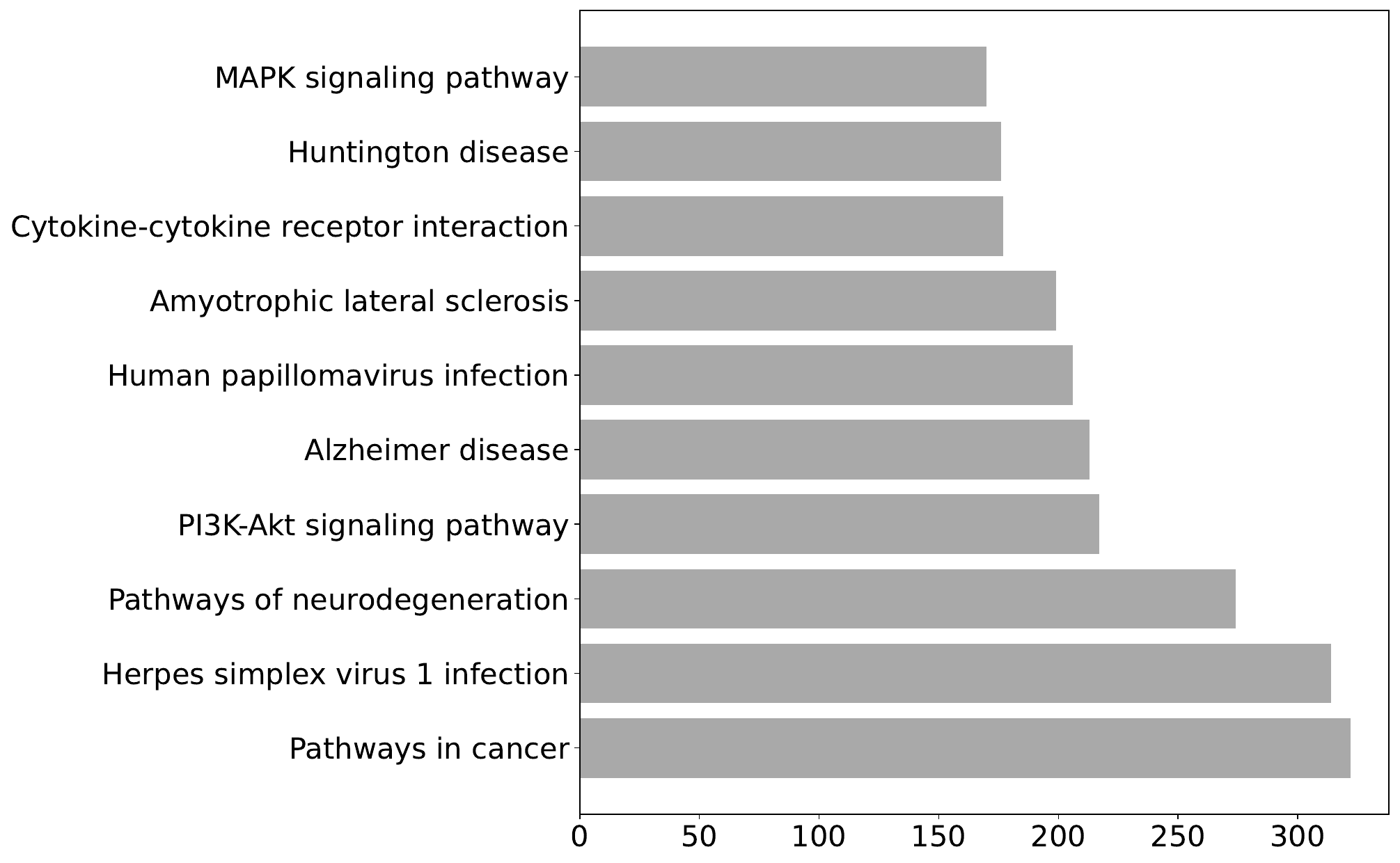}
\caption{Resuls of KEGG pathway analysis of all up- and down-regulated genes in host viral response, visualization of 10 most common KEGG pathways for DEGs. The presented results were calculated for the GSE80697 dataset.}
\label{fig:mpr3}
\end{figure}

The expression of various cell systems is altered by viral infections. Thus, we conduct an analysis of differentially DEGs for several datasets (GSE80059, GSE76600, GSE80697, GSE45399, GSE79458, GSE86528, and GSE48142). After quantile normalization and preprocessing for all datasets, we performed DEG analysis between two groups: virus-treated and mock-treated cells (Fig.\ref{fig:mpr1}). All DEGs were then selected for Gene Ontology (GO) and KEGG pathway analysis (Fig.\ref{fig:mpr2}, Fig.~\ref{fig:mpr3}).

The DEG analysis for the Ebola dataset revealed top-ranking genes by $|\log_2 FC|$ that have previously been identified as potential biomarkers or DEGs for viral infections. These include hsa-miR-3676-5p \cite{Brameier2013}, hsa-miR-4443 \cite{Xun2015}, hsa-miR-762 \cite{Yoneyama2021}, hsa-miR-4428 \cite{Luo2020}, hsa-miR-4446-3p \cite{Ghosh2022}, hsa-miR-494 \cite{Chen2022}, hsa-miR-210 \cite{Russo2011}, and miRNAs such as hsa-miR-21-3p \cite{Luo2022, Amirfallah2021}, hsa-miR-483-3p \cite{Maemura2018, Yue2018}, and hsa-miR-197-3p \cite{Chen2013}, which have experimental data confirming their possible roles in regulating host viral responses.

For the influenza virus datasets (GSE76600, GSE80697, GSE45399), we found that top-ranking genes by $|\log_2 FC|$ were involved in interferon (IFN)-mediated antiviral responses, including IL8, IL6, IFIT2, IFIT1, IFIT3, IFITM1, and the innate sensor ZBP1 \cite{Kuriakose2018}, GPI-linked cell surface receptor FOLR2 \cite{Gupta2022}, and interferon-inducible antiviral proteins RSAD2 \cite{Qiao2022}, MX1 \cite{Busnadiego2014}, and CXCL10 \cite{Huang2020, Pugh2022}. To explore the functions of these genes, we used Gene Set Enrichment Analysis (GSEA) to analyze the enrichment of genes in biological processes (BP), setting the false discovery rate (FDR) at $p < 0.05$ as the selection criterion. GSEA showed that the identified DEGs were significantly enriched in processes such as the cellular response to type I interferon, type I interferon signaling pathway, glucan biosynthetic process, glycogen biosynthetic process, and the negative regulation of viral processes. Additionally, interferon-related long non-coding RNAs (lncRNAs) were identified in our analysis of top-ranking genes, including small NF90 (ILF3)-associated RNA A3 (SNAR-A3) and IFNA22P, an interferon alpha 22 pseudogene. For GSE76600, we observed significant differences in the expression levels of miRNAs involved in various cellular processes, including cell migration and proliferation, such as hsa-miR-122-5p \cite{miWang2019, Raitoharju2016, Dai2020}, hsa-miR-100-3p \cite{Peng2019}, hsa-miR-29b-1-5p \cite{Kim2020}, hsa-miR-23b-5p \cite{Xian2018}, hsa-miR-222-5p \cite{Song2019}, hsa-miR-141-5p \cite{Bao2019}, and miRNAs involved in inflammation processes, identified as biomarkers for viral infections, such as hsa-miR-200c-5p \cite{Abdolahi2022}, hsa-miR-21-3p \cite{Luo2022, Amirfallah2021}, hsa-miR-3591-3p \cite{Liu2019}, and hsa-miR-15b-3p \cite{Huang2022}.

For the MERS-coronavirus datasets (GSE79458, GSE86528), we identified top-ranking genes by $|\log_2 FC|$ that included transcription repressor SSX2 \cite{Jha2020}, integral membrane component TSPAN18 \cite{Myall2021, Paterson2021}, receptors such as OR13D1 and OR3A4P, transcription factors NFE4 \cite{Tabone2018}, SOX13 \cite{Wang2021}, and LCE2B \cite{Huang2020}, ion transport regulator FXYD3 \cite{Lavorgna2021}, regulator of cytoskeletal organization and signal transduction RHOF \cite{Shaverdashvili2019, Tian2023}, mitochondrial antiviral signaling protein HRK \cite{Boehler2019}, and ABL2 kinase, which prevents the virus from entering the body \cite{Liu2021, Naik2022}. GSEA confirmed that the identified DEGs were significantly enriched in processes such as SRP-dependent cotranslational protein targeting the membrane, cotranslational protein targeting to the membrane, cytoplasmic translation, cellular protein metabolic processes, stress granule assembly, regulation of the apoptotic process in vascular smooth muscle cells, negative regulation of glucocorticoid receptor signaling, and cardiac muscle cell action potential, as well as mitochondrial ATP synthesis coupled to electron transport.

For the SARS-coronavirus dataset (GSE48142), we found that top-ranking genes by $|\log_2 FC|$ involved in interferon (IFN)-mediated antiviral responses included IFNL3~\cite{O’Connor2014}, IFNB1~\cite{Bourdon2020}, IFNL1~\cite{Hemann2017}, IL28B~\cite{Chinnaswamy2014}, IL29 ~cite{Kelm2016}, antiviral factors CREB5~\cite{Zhang2016}, and TRIM22~\cite{Pagani2021}, along with the transmembrane receptor RGR~\cite{Hooks2008} and chemokine CXCL10~\cite{Huang2020, PPR:PPR571138}, which play a role in IFN-beta immuno-protection in the retina. The results of GSEA revealed that DEGs were enriched in processes such as cotranslational protein targeting to the membrane, SRP-dependent cotranslational protein targeting to the membrane, protein targeting to the ER, cellular response to type I interferon, interferon-gamma-mediated signaling, regulation of viral genome replication, and cytokine-mediated signaling pathways.

GO and KEGG analysis revealed that most of the genes are involved in processes such as transcription regulation, gene expression regulation, apoptotic process regulation, and cytokine-mediated signaling pathways (including MAPK and PI3K-Akt pathways), as well as pathways related to cancer and infectious diseases. These data confirm the role of the identified genes in regulating the viral life cycle and viral genome replication.

Overall, we identified genes involved in the interferon-mediated antiviral response, as well as genes critical in detecting and responding to viral infections, such as innate sensors and receptors, transcription regulators, and mitochondrial antiviral signaling proteins. However, we also identified some genes that have not been extensively studied in the context of antiviral processes. Some studies suggest that these genes may play a role in modulating the host immune response to viral infections, and further studies are needed to elucidate their specific roles and mechanisms of action in antiviral processes.


\subsection{Construction TOM and following SimCIM analysis}
\begin{figure}[h!]
\centering
\includegraphics[width=1\linewidth]{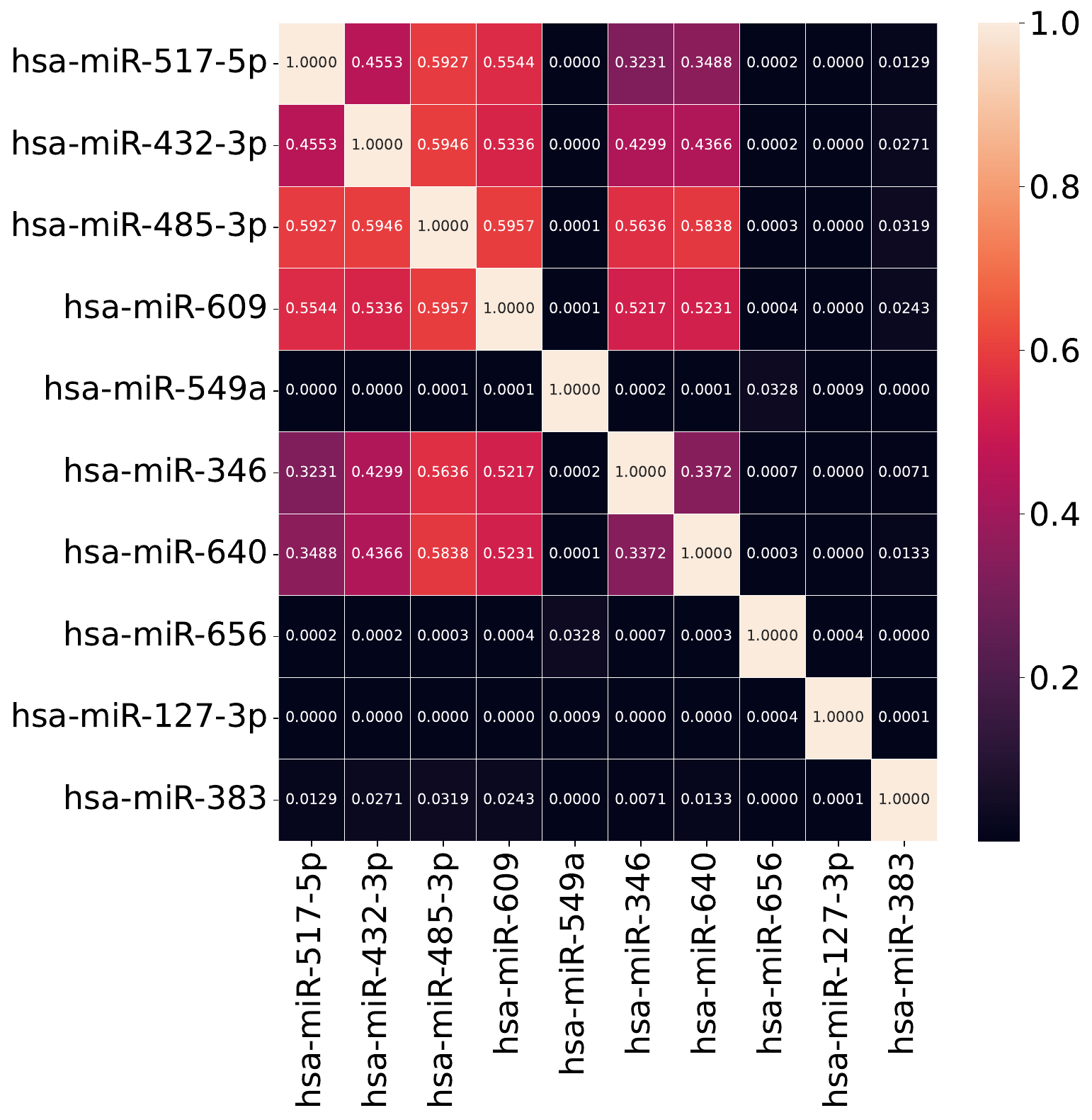}
\caption{TOM visualization for ten genes of GSE80059 dataset. TOM calculated following the standard methods in PyWGCNA.}
\label{fig:mpr4}
\end{figure}

Another large-scale genome-wide association approach for identifying significant genes involved in biological processes is weighted correlation network analysis (WGCNA). This method uncovers hidden relationships between genes exhibiting similar patterns of expression changes and disease traits. The WGCNA tool was used to construct a TOM (Fig.\ref{fig:mpr4}), specifically identifying highly correlated genes to provide unbiased targets.

\subsection{SimCIM result verification}
We applied a quantum-inspired algorithm to identify key genes in the in vitro model of viral response. The algorithm enabled us to pinpoint the most relevant genes for each selected GSE dataset.

To validate these gene sets, we examined publicly available data to identify potential molecular partners and cellular processes involving these target genes. We analyzed SPLASH and CLIP-seq data for miRNA–mRNA interactions available in Mirnet, and explored lncRNA-RNA and RNA-RNA interactions in the RISE database (a database of RNA interactomes).

For the Ebola virus dataset, we identified a set of miRNAs involved in pathological processes and viral response regulation, such as hsa-miR-33b-3p \cite{He2017}, hsa-let-7b-3p \cite{Letafati2022}, hsa-miR-4709-3p \cite{Jiang2021}, and hsa-miR-640 \cite{Harel2020, Zhai2019}, among others. We also found interactions with immune response and cell cycle regulation mRNAs like ATP1B3 \cite{Lu2016}, CSK \cite{Gao2020}, and HMGB1 \cite{Ding2021}.

For the influenza virus dataset, we identified mRNA and lncRNA genes such as FAM96A \cite{Ebrahimi2022}, C1orf131 \cite{Replogle2022}, and IL15 \cite{Verbist2012}, involved in host viral response. The algorithm also highlighted miRNAs such as hsa-miR-4750-5p \cite{Pollet2023}, hsa-miR-4734 \cite{Egaña-Gorroño2016}, and hsa-miR-122-5p \cite{Saha2022} as key players in viral infection pathophysiology. Additionally, miRNA–mRNA interactions with hsa-let-7b-5p \cite{Letafati2022} and hsa-mir-30a-5p \cite{Poore2018} were identified, emphasizing their role in regulating inflammatory protein activity.

For MERS-CoV, we found genes playing critical roles in regulating transcription and immune response, such as PGK1 \cite{Li2018}, HAX1 \cite{Balcerak2022}, and PRDM4 \cite{He2016}. Pathway analysis revealed key genes involved in the response to viral infections, including C14orf166 \cite{Rodriguez-Frandsen2016}, CD302 \cite{Reinecke2022}, and RCBTB2 \cite{Saayman2016}. We also identified potential lncRNA interactions, with TINCR, NEAT1 \cite{Wang2017, Yang2022} playing a significant role in the host antiviral response.

For SARS-CoV-2, quantum-inspired analysis revealed genes involved in basic cellular processes and immune responses, such as PLK1 \cite{Biswas2021}, RDM1 \cite{Hamimes2006}, and IFIT-1 \cite{Lozhkov2022}. Analysis of miRNA–mRNA interactions indicated key miRNAs involved in metabolism and inflammation regulation, including hsa-let-7b-5p, hsa-let-7e-5p, and hsa-mir-15a-5p. RNA interactome analysis further revealed interactions with lncRNAs like TINCR and LINC01288 \cite{Bian2019}, LINC00482 \cite{Liu2020}.

Interestingly, SARS-CoV-2 and MERS-CoV share genomic similarities, with approximately 50\% overlap. The predicted miRNA sets for these viruses overlapped by over 60\%, with 64 common miRNAs identified for both.

\section{Discussion}
The rapid growth of genome-wide screening technologies has enabled advances in drug development, functional human genomics, and basic science. These technologies have led to the transformation of modern biology and require new data analysis methods. Our work highlights two methods for identifying significant genes involved in a viral response. We provide DEG analysis and an analysis using a quantum-inspired algorithm for seven microarray datasets.

DEG analysis is one of the most popular tools for identifying significant genes. However, this method has some limitations. It has been shown that the general agreement regarding the identification of DEGs across all tools is low \cite{DEG_Wang2019}. When comparing the top 1000 DE genes identified by each tool, only 92 common DEGs were detected across all tools. Therefore, there is a need to develop alternative approaches for genome-wide data analysis.

Viral infection induces massive changes in the host transcriptome. These gene expression changes include the regulation of metabolic aberrations and the modulation of the immune response. Using expression data from microarray analysis, we performed DEG analysis to better understand the molecular basis of viral infection and identify potential biomarkers. We found genes that were differentially expressed ($p < 0.05$) and counted $|\log_2 FC|$. We compared viral-treated to mock-treated controls. GO functional and KEGG pathway enrichment analysis revealed that these genes belong to the interferon signaling pathway, including IL8, IL6, IFIT2, IFIT1, IFIT3, IFITM1, IL28B, and IL29; a regulator of leukocyte chemotaxis, CXCL10; a regulator of metabolic reprogramming of cells, RSAD2; and transmembrane receptors such as FOLR2, OR13D1, OR3A4P, SLC22A2, TTLL6. Other genes involved in regulating transcription, gene expression, apoptosis, and various signaling pathways (cytokine-mediated, MAPK, PI3K-Akt) also play significant roles in restricting viral infection. We also detected lncRNAs involved in the interferon pathway and miRNAs reported to have different expression levels in viral infections \cite{Huang2018}.

As mentioned above, there is low agreement between the results of DEG determination methods. Analysis using a quantum-inspired algorithm presents an alternative approach for identifying significant genes involved in the host immune response. The considered approach helps detect complex sets of interactions. We identified genes involved in essential cellular processes such as DNA damage response (HMGB1, WNT2B), DNA replication (PRIM1), transcription initiation (MED20), initiation of translation (RPL30, RPS4X), mRNA processing (PCBP2), the cell cycle (PLK1), and proteasome degradation (PSME1). We also found metabolic regulators such as those involved in vitamin B12 metabolism (SOD), energy metabolism (PPR1), the vitamin K cycle (VKORC1), and fatty acid beta-oxidation (ECI1). Moreover, we detected genes that play crucial roles in different signaling pathways, including type II interferon signaling (ICAM1), Aurora A or B signaling (TDRD7, PEPB1), VEGFA-VEGFR2 signaling (SLC8A1, PGK1), chemokine signaling (CSK, PRKX), Notch signaling (MAML3), and PI3K-Akt signaling (CREB3, EFNA2, RAB2A, EFNA2), among others. Additionally, our data showed that these genes correspond to the inflammation process—such as the Wnt pathway (SERPINF1), p53 pathway (CSE1L), T cell activation (PCBP2), macrophage markers (CD14), and cytokine inflammatory response (IL15). Notably, we observed genes (CD274, PGK1, ICAM1) in the CPTAC therapeutics database as possible drug targets.

The observed microarray datasets include genes from different biotypes—mRNA, miRNA, and lncRNA. Several lncRNAs and miRNAs were discovered to regulate gene expression and influence the pathogenesis of various human diseases. Cross-linked ncRNAs and mRNAs formed multilayered regulatory networks. Analysis of possible ncRNA and mRNA interactions provides new insights into the host viral response regulatory mechanisms. miRNAs are small ncRNA molecules, approximately 22 bases in length. They negatively regulate gene expression at the post-transcriptional level by binding to their target mRNAs. A single miRNA can regulate multiple target RNAs. It has been indicated that about 30$\%$ of human mRNAs are regulated by miRNAs \cite{Qi2015}. Several studies have shown that miRNAs can regulate the antiviral response by targeting different genes involved in the pathways and directly targeting viral RNA molecules. miRNAs lay the foundation for constructing a ceRNA network. On the other hand, long non-coding RNAs (lncRNAs) are a subset of non-coding RNAs longer than 200 nucleotides. Host lncRNAs have been reported to regulate the innate antiviral response and may inhibit or facilitate viral infection \cite{Wang2020, Liu2017}. Considering the significance of ncRNAs in viral infection progression, quantum algorithm analysis was followed by searching for possible lncRNA-mRNA and mRNA-miRNA interactions.

Although the mechanisms underlying changes in viral pathogenesis are not fully understood, some partial mechanisms and target genes have been reported. The results of the miRNA-mRNA prediction help identify a set of significant miRNAs. Most genes in these miRNA sets are associated with the host antiviral response. We also considered a few miRNAs with well-understood mechanisms of action. Thus, we identified miR-16-5p, miR-21, and miRNAs from the miR-34/449 family, which induce cell cycle arrest and apoptosis regulation. For all datasets, we found miRNAs from the miR-let-7 family. It has been shown that let-7 may affect the antiviral response of the human immune system by reducing the expression levels of several target genes (MAP4K4, STAT3, IL-10, ORF1ab, SARS-CoV-2, H1N1 M1, etc.). For miR-106b-5p, miR-20b-5p, miR-342–3p, miR-320d, and miR-92b-5p, it is suggested that they participate in the activation pathways of TGF-$\beta$, insulin, and T and B cell receptors \cite{ShWang2017}. Several identified miRNAs may target viral sequences or cellular genes associated with the viral life cycle, such as interactions with nucleocapsid or spike proteins (has-mir-103a-3p, hsa-mir-29a-3p, hsa-mir-29c-3p, hsa-mir-103a-3p) and cellular factors (hsa-mir-1-3p, hsa-let-7b, hsa-mir-34a-5p). Additionally, analysis of lncRNA-mRNA interactions revealed that TINCR and NEAT1 play important roles in transcription regulation and gene expression during viral infections. The mRNA-miRNA and mRNA-lncRNA analyses confirmed the significance of gene sets detected by the quantum algorithm.

Drawing from the results of the comparison of the human host response to various viral infections, as determined by the SimCIM and DEG methods, it can be inferred that both techniques successfully pinpoint genes involved in the regulation of gene expression and protein processing. Similar roles of the identified genes in the regulation of metabolism and activation of signaling pathways can also be noted. However, there were differences. For instance, DEG analysis revealed genes, most of which are involved in the interferon response. Genes identified in the SimCIM analysis cover a broader range of viral response signaling pathways. Additionally, SimCIM identified genes (such as IL15) that are specifically associated with coronavirus infections and helped pinpoint specific disruptions in metabolic pathways. Thus, in summary, we can conclude that SimCIM can be used as an alternative to DEG for identifying key genes in the viral response. This conclusion leads us to consider the effectiveness of the SimCIM method for microarray data analysis. Furthermore, the key genes identified, along with their molecular partners, form a solid foundation for subsequent validation studies in cell cultures and animal models, which could prove instrumental in the development of future host-targeted drugs.

\section{Acknowledgment}
This work was supported by Russian Science Foundation (19-71-10092). Authors also acknowledge the support from Rosatom in the framework of the Roadmap for Quantum computing (Contract No. 868-1.3-15/15-2021 dated October 5, 2021).

\begin{figure}[h]
\centering
\includegraphics[width=1\linewidth]{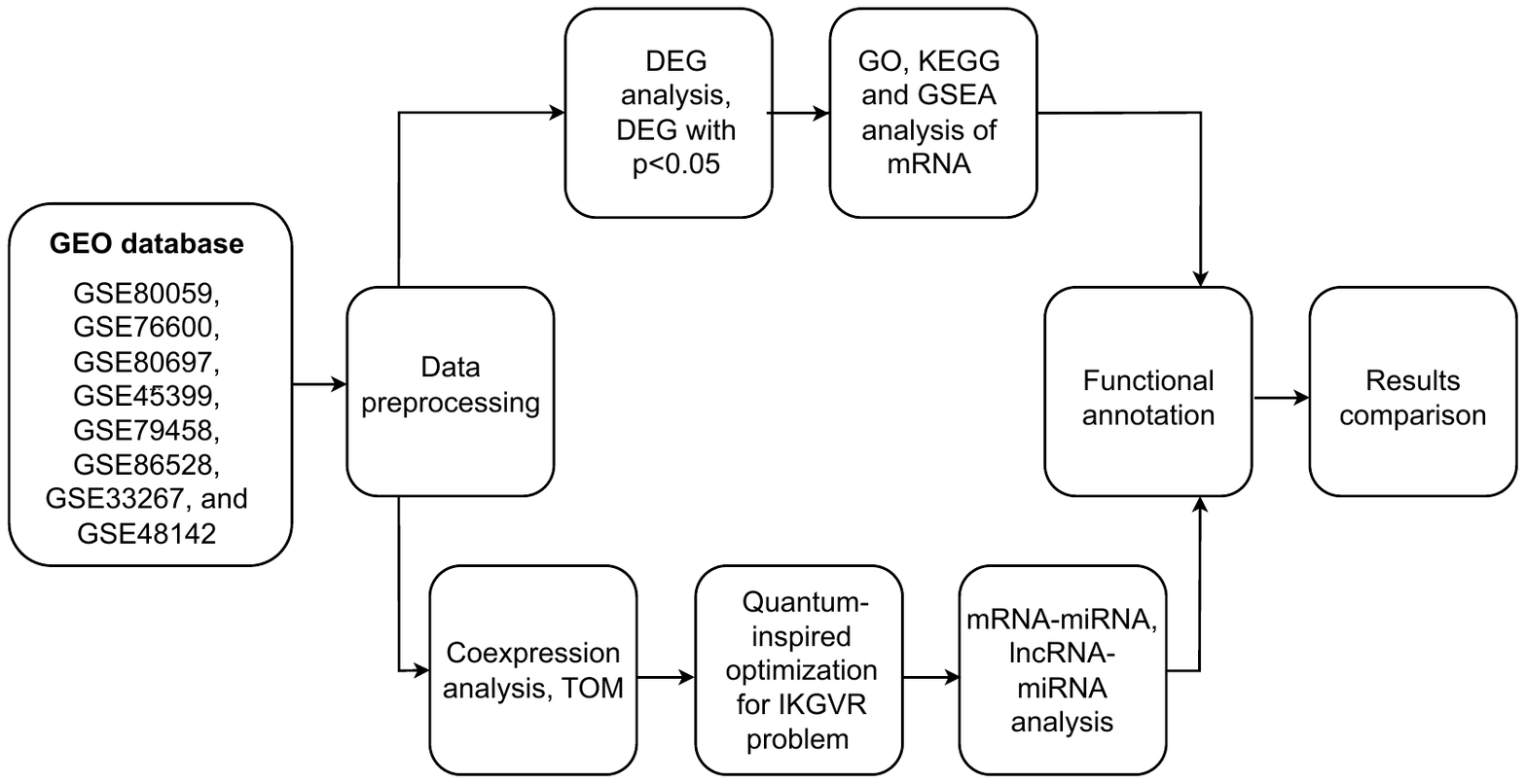}
\label{fig:mpr5}
\end{figure}

\bibliographystyle{ieeetr}

\bibliography{references}

\end{document}